# Duration of interactions in quantum electrodynamics: basic concepts, temporal features of kinetic phenomena

Mark E. Perel'man[1)]
*Racah Institute of Physics, Hebrew University, Jerusalem, Israel*

**Abstract**  For temporal magnitudes describing, in details, processes of particles scattering for a long time at each necessity case a particular, ad hoc reception were used. However the desirability of general approach basing on concepts of quantum field theory and uniting two main problems, durations of delay at scattering and of formation of arisen states, seems evident.

We show that the Ward-Takahashi identity allows the defining of general temporal function in the frame of QED, real part of which describes the duration of delay of scattered particle on scatterer and imaginary part describes the duration of formation ("dressing") of outgoing bare particle. The functions of same form can be revealed in the known QED descriptions of processes, i.e. they arise in the course of standard calculations, but are not entered artificially. These functions submit to the operator of duration determinable by the reciprocal analogue of the Schrödinger equation, operator of which is canonically conjugated with Hamiltonian. Definition of temporal functions leads to the developing of such kinetic theories: theory of optical dispersion (with the decision of certain known paradoxes), theory of multiphoton processes (definitions of their thresholds and saturations) and theory of phase transitions (determination of general correlation radii and the full system of critical indices). The determination of temporal functions demonstrates also the instanton-type nature of tunneling processes, makes clear sense of the adiabatic hypothesis of quantum theory and some renormalization schemes. In an application to the problem of charge nullification they lead to a restriction of allowable number of fermions of the theory.

**1. Introduction**

Concepts of duration of the elementary scattering act, i.e. temporary delays of scattered particles by scatterers and durations of terminating of particles or states formation, for a long time remained unclaimed in quantum physics. However, it is necessary to emphasize, similar problems long since became essential and were investigating in the problems of telegraphy (signals delay and expand into long communication lines, e.g. [1]).

The concept of duration of physical state (e.g. photon) formation has appeared, as far as I know, with the beginning of research of Ĉerenkov radiation. From the classical point of view an electron with the velocity bigger the light speed in medium must continuously be emitting energy. But as the observed electrons are emitting photons with the concrete parameters periodically, it means, as Frank had shown [2], that for *formation* of every photon or for *retuning and restoration* of electron as emitter the definite time (or path length) of formation is needed:

$$\tau_2(\omega) \approx \pi/\omega|1 - \beta n\cos\theta| = \pi/|\omega_0 - \omega|, \qquad (1.1)$$

---

[1]). E-mails: mark_perelman@mail.ru; m.e.perelman@gmail.com .

where ω is the frequency of emitted photon, θ is the Čerenkov angle, β = v/c, n is the index of refraction, $\omega_0 = \vec{k} \cdot \vec{v}$. This estimation had been calculated pure classically, via the Huygens principle; but completely coincides with the QED result (we shall return to it below).

Similar effects are known in the theory of electron bremsstrahlung in the nuclear field: electron is accumulating energy on the definite length of path [3], i.e. during the definite time. Here, just as above, must be marked, that this time duration can be related also with the retuning of emitting electron, its "relaxation" after one act of emission or absorption till the subsequent one.

There are also several other phenomena, which could be considered via existence of definite path (or time) of formation: synchrotron radiation, transient Ginzburg-Frank radiation and so on [4]. And must be specially remarked that certain these phenomena can be deduced by pure classical consideration, by the Huygens principle, and these values become usually bigger the quantum uncertainty. It shows that the duration of interactions, in general, can not be a simple corollary following uncertainty relations.

On the other hand as the pure quantum phenomena are presented such effects. The cloud of virtual formations must surround each physical particle, but newborn particles are bare and therefore some temporal duration for their dressing is needed. Such phenomena were sometimes observed in the cosmic rays investigations, when particle of very high-energy flews after interaction act along some path without interactions. It should be interpreted as the time period needed for restoring particles state, its dressing, i.e. formation of its physical coat, the cloud of virtual particles [5]. By that time Moshinsky had shown that the transition of atomic electron on upper state must pass via the "time diffraction" period, when electron's energy oscillates with damping around stationary value [6].

Temporal characteristics of scattering process should include, in principle, two types of magnitudes: duration of a delay of colliding particles during their interaction (their temporary, i.e. virtual association) and duration of formation (dressing) of products of reaction, sometimes after tunneling. Such researches have a long history: in the beginning of development of quantum theory McColl had shown that calculation of duration of tunnel transition conducts to negative value [7]. Therefore during long time was factually recognized that any estimations of temporal characteristics, besides the uncertainty principles, are practically not necessary or even impossible.

The first (semi-qualitative) consideration of time delay in processes of tunneling had been performed, as far as I know, by Bohm [8]. The more constructive and physically more transparent magnitude of time delay under an elastic scattering was introduced by Wigner [9] through the partial phase shifts, $\tau_l(\omega) = d\delta_l / d\omega$, generalized by Smith [10] via *S*-matrix as

$$\tau_l(\omega) = \text{Re}(\partial / i\partial\omega) \ln S = (\partial / i\partial\omega) \arg S. \qquad (1.2)$$

Then Goldberger and Watson had deduced on the base of (1.2) a "coarse-grain" Schrödinger equation, by which the generality of this definition had been shown [11]. But at their approach the magnitude (1.2) had been introduced artificially, by the serial decomposition of Fourier transformed response function *S(t)* of linear system or its logarithm near the selected frequency without discussion of its imaginary part, higher terms and dependence on space variables.

Another approach, which seems at first glance distinctive from the Wigner–Smith one, was suggested by Baz' [12] for consideration of nonrelativistic tunneling processes: to a scattering particle is attributed magnetic moment and its rotation at the scattering process is analyzed (the method of "Larmor clocks").

In the rather investigations Pollak and Miller [13] had introduced another magnitude:

$$\tau_2(\omega) = \mathrm{Im}(\partial/i\partial\omega)\ln S = (\partial/i\partial\omega)\ln|S|, \tag{1.3}$$

as the definite duration of tunneling, but it can be considered more widely as the duration of state formation, its "dressing". (Note that the duration of particle formatting can be calculated by dynamical considerations, it presents the special direction in the high energy physics; cf. the review [14].)

After these initial investigations a number of various definitions of duration of scattering processes and interaction were offered, different determinations of duration of interactions are introduced, e.g. the reviews collected in [15].

With all these directions certain principal problems must be naturally posed:

1). Are in general significant and observable and in which limits the temporary sequences of the examination of quantum processes? Or the limitation of evaluations of complete duration by the uncertainty principles, in a spirit of the initial Heisenberg concept of S-matrix, is not only possible, but also is enough?

2). Are needed an introduction of some new definitions of duration of processes or they are yet contained in the theory in certain non-evident form and can be extracted from existing representations? (The excellent accuracy of the QED calculations may mean that if such magnitudes are not contained within its frame, their additional, ad hoc, depositing hardly can be justified.)

3). Are there such phenomena, explanations of which are yet impossible beyond the explicit reference onto the theory of temporal functions (the general Ockham principle)?

4). Can the investigation of temporary functions be resulting in anything definitely new within our understanding of quantum phenomena? And also, more generally: are needed certain additions or corrections of common theory?

The answer onto the first question can be presented as well known, e.g. by the investigations of Ĉerenkov radiation and bremsstrahlung. On the other hand, possibility of corresponding calculations in the frame of theories, long since known, hampers the recognition of their principal novelty or the desirability of special temporal concepts.

The answer onto the second question can be given by some our articles. The temporal magnitudes as manifestation of features of corresponding functions can be found out in relativistic dispersion relations [16]; they are naturally appear in the radius of convergence at summation of complete perturbative series of multiphoton processes as certain thresholds (as the opportunity of capture of the following photon is determined by the duration of virtual keeping of previously captured energy by scatterer, they determine opening of new channels of processes [17]; more completely they are considered in [18], the Section 5 below). Further has been revealed that temporal functions are directly connected with propagators of particles and it explains why the calculations without their direct introduction are possible [19]. Thereby these functions are not entering artificially, a priori; a more systematic investigation of existing theories reveals their presence in calculated expressions and their physical significance.

For a partial answer onto the third question such results can be mentioned. The revealing of temporal magnitudes at calculation of multiphoton processes shows phenomena of saturation and determines the thresholds of new channels of reactions [17]. The application of these methods to the heat field allows the consideration of problems of phase transitions: the removal of latent heat at the transition of the first kind into more ordered states by characteristic radiation [20] (these new phenomena were investigated and repeatedly experimentally proved, see [21] and references therein). Determination on this base of the complete system of critical indices at phase transitions and the proving of several previous assumptions in this field became possible [22] (Section 6 below).

Another field, for which the application of temporal functions represents the most useful, is the theory of optical dispersion. Temporal functions naturally allow consideration of photons kinetics at their pass in substance, i.e. the calculation of their group velocity and consequently the indices of refraction [23]. All it leads to decision of certain old paradoxes [24] (Section 4 below).

The fourth question is more complicated and answers depend on a scientific taste, not only from concrete results. Our answer consists in establishing of general form of equation for temporal functions that unites (1.2) and (1.3) as

$$\frac{\partial}{i\partial\omega}S(\omega) = \hat{\tau}\cdot S(\omega),  \qquad (1.4)$$

i.e. as the reciprocal Schrödinger equation with a temporal operator instead the Hamiltonian [25] (reciprocal Klein-Gordon equation was mentioned in [16]), this equation can be considered as the basic one for several different kinetic problems. On the other hand with this question can be related possibilities of some restricted superluminal phenomena and "nonlocality in the small" first considered in [26].

Thus we can *assert* that the conception of temporal functions should, besides its general cognitive significance, give new comprehension of some old problems and should represent the basis for detailed description of several kinetic phenomena.

Our consideration will follow in such sequence. In the Section 2 the general temporal functions are presented; their deductions in the frame of QED are given in the Section 3 (it allows, in particular, consideration of such phenomenon as zitterbewegung). In the Sections 4-6 some above cited concrete applications are examined that, in particular, results in explaining of certain known paradoxes. Section 7 is devoted to problems of tunneling. In the Section 8 the justifying of adiabatic hypothesis and its connection with temporal functions is examined; this line is continued in the subsequent section relative to renormalization procedures. In the Section 10 certain high energy features of Green functions are briefly considered. The main results are summed in the Conclusions.

## 2. Photon Green functions and durations

Temporal functions are introducing in general form via the response functions or matrix elements of scattering processes as transfer functions in the frequency's representation, $S(\omega) = |S(\omega)|\exp(i\Phi(\omega))$, other variables are omitted:

$$\tau(\omega) \equiv \tau_1 + i\tau_2 = (d/id\omega)\ln S(\omega).  \qquad (2.1)$$

In this expression

$$\tau_1 = \operatorname{Re}\tau(\omega) = d\Phi/d\omega \quad \text{and} \quad \tau_2(\omega) \equiv \operatorname{Im}\tau(\omega) = (d/d\omega)\ln|S(\omega)|  \qquad (2.2)$$

describe, correspondingly, the delay duration at the scattering process via the variation of response phase and the duration of outgoing particle formation (its "dressing" or "redress") via the amplitude alteration [25].

As the simplest illustrative examples of temporal functions, the propagators in the nearest order in the $(\omega,k)$ representation could be considered. So the causal photon propagator of the lowest order in the Feynman gauge, $D_c(\omega,\vec{k}) = 4\pi/(\omega^2 - c^2 k^2 - i\eta), \quad \eta \to 0+$, leads to the temporal functions:

$$\tau_1 = 2\pi\omega\cdot\delta(\omega^2 - c^2 k^2) = \pi(\delta(\omega - c|k|) + \delta(\omega + c|k|));$$

$$\tau_2 = 2\omega/(\omega^2 - c^2k^2) = \left(1/(\omega - c|k|) + 1/(\omega + c|k|)\right). \tag{2.3}$$

The time delay $\tau_1$ is non-zero for real photons only ($\omega^2 = c^2k^2$) and for the static Coulomb field ($\omega=k=0$). The duration of dressing $\tau_2$ is finite for virtual photons with $\omega^2 \neq c^2k^2$, its negative sign at $\omega < c|k|$ corresponds to tunneling processes [27] (the case of anomalous dispersion as example). The conformity of (1.1) with (2.3) is evident.

The interpretation of temporal functions can be achieved also by the presentation of temporal functions (2.3) in x-space via the Green functions of Klein-Gordon equation:

$$\tau_1(x) = i\partial_t(D_{ret}(x) + D_{adv}(x)); \quad \tau_2(x) = i\partial_t\left(D^{(+)}(x) - D^{(-)}(x)\right) \tag{2.4}$$

These relations present the delay duration at scattering process as the jointed decay of retarded and generation of advanced waves or vice versa in the light cone limits. But the duration of formation (dressing) shows that the physical particle formation requires waves with different signs of energy, i.e. vacuum fluctuations including out-of-cone ones are take part in them, therefore superluminal effects in these processes cannot be excluded. (These examinations are shown, in particular, that superluminal phenomena should be present and, in principle, may be observed at all scattering processes, not only in the QED.)

Temporary functions allow certain specification of properties of the basic singular functions of field theory. Let us determine their corresponding temporal functions in the $\omega$–$r$ representation ($\hbar = c = 1$ below):

$$\tau(D_{R,A} \mid \omega, \vec{r}) = \pm r; \quad \tau(D_c \mid \omega, \vec{r}) = r\,\mathrm{sgn}(\omega); \tag{2.5}$$

$$\tau(D, D_1, D^{(\pm)} \mid \omega, \vec{r}) = -ir\cot(\omega r); \tag{2.6}$$

$$\tau(\overline{D} \mid \omega, \vec{r}) = ir\tan(\omega r). \tag{2.7}$$

The representations (2.5) demonstrate that propagators of "dressed" photons can be considered classically, they describe passage of photons without any reformations.

The representations (2.6) show that commutations of field intensities described by the Pauli-Jordan function are connected with their formation or reformations without delay. The Coulomb field is infinitely in the undressed state and therefore the term $1/\omega$ leading to the Coulomb pole in (2.6) must be subtracted. The subtraction can be performed with the decomposition of cotangent: $\cot(x) = 1/x + 2x\sum_1^\infty 1/(x^2 - \pi^2 n^2)$. It leads to the renormalized expression for (2.6):

$$\tau_2^{(renorm)}(\omega, \vec{r}) = -2\omega r\sum_1^\infty 1/(\omega^2 r^2 - \pi^2 n^2), \tag{2.8}$$

which shows that the first pole of (2.6) is at the point $\omega r = \pi$. Therefore it can be taken that the formation path for photon is of the order:

$$\Delta l \sim \pi/\omega = \lambda/2. \tag{2.9}$$

As this process is instantaneous, it corresponds to the jump of photon at the act of formation on the distance $\lambda/2$ (cf. [27], where this estimation is proven by the covariant dispersion relations). Really it means impossibility of gradual photons formation that completely corresponds to quantum paradigm.

As must be underlined, (2.9) and (2.3) strictly correspond to duration of Ĉerenkov photon formation (1.1) that justifies results of classical approach to this phenomenon.

For analysis of near field the usual decomposition can be written out (e.g. [28], compare [29, 30], certain problems of near field are considered in [31]):

$$D_{ij}(\omega,\vec{r}) = \left\{(\delta_{ij} + e_i e_j) - \frac{i}{\omega r} P_{ij} \cot(\omega r) + \frac{1}{(\omega r)^2} P_{ij}\right\} D(\omega,\vec{r}) \quad (2.10)$$

with directing cosinuses $e_i = x_i/r$ and the tensor $P_{ij} = \delta_{ij} - 3e_i e_j$. Three terms of (2.10) are related, correspondingly, to far, intermediate and near fields. The Sommerfeld condition of radiation, $(\vec{r} \cdot i\vec{\nabla}_r)U(\vec{r}) = |\vec{k}||\vec{r}|U(\vec{r})$ at $r \to \infty$, selects within these terms the far field only.

Near field part corresponds to the function entered by Schwinger in [32]:

$$D_N(\omega,\vec{r}) = -(1/2\pi\omega^2 r^3)\sin(\omega r). \quad (2.11)$$

Corresponding duration,

$$\tau(D_N | \omega,\vec{r}) = i\left(\frac{2}{\omega} - r\cot(\omega r)\right), \quad (2.12)$$

shows that formation of near field requires duration 4-times bigger the uncertainty value in addition to (2.12). Therefore this value can be measurable.

Note that intermediate and near fields have space-like parts, corresponding instantaneous jumps (2.9), the singular function (2.11) describes connection of atoms in near field and transferring of excitations that can be instantaneous [31].

It can be noted that the spontaneous breaking of symmetry that leads to phase transitions of the first kind into more ordered states, at least, is executed by emission of photons [21], i.e. it corresponds to the Higgs mode.

Electromagnetic interactions in media can be examined via the dielectric susceptibility:

$$\varepsilon(\omega;\vec{r}_1,\vec{r}_2) = f(\vec{r}_1,\vec{r}_2)/i[(\omega_0 + i\Gamma/2)^2 - \omega^2] \quad (2.13)$$

(we consider for brevity the two-level system only), which can be considered as the response function and for which temporal functions can be expressed just as above, but in the ($\omega$-$r$) representation. Thus the logarithmic derivative of (2.13) leads to

$$\tau_1(\omega) = \frac{\Gamma/2}{(\omega-\omega_0)^2 + \Gamma^2/4}; \quad \tau_2(\omega) = \frac{\omega-\omega_0}{(\omega-\omega_0)^2 + \Gamma^2/4}. \quad (2.14)$$

These forms will be used below.

## 3. Ward identity, proper duration of interaction, zitterbewegung

Although several different possibilities for establishment of temporal functions had been considered in [25], it seems that the most short and the closest one to canonical theory follow the Ward identity.

Let's consider the simplest Ward-Takahashi identity:

$$k_\mu \Gamma_\mu(p,q,k) = G^{-1}(p) - G^{-1}(q), \quad (3.1)$$

where $\Gamma_\mu$ is the vertex part and $G(p)$ is the causal propagator. At $q \to p$ its right part can be transformed as

$$k_\mu \frac{\partial}{\partial k_\mu} G^{-1} = -iG^{-1} k_\mu \frac{\partial}{i\partial k_\mu} \ln G = -iG^{-1} k_\mu \xi_\mu, \quad (3.2)$$

i.e. can be rewritten via the covariant temporal function, corresponding to 4D generalization of (2.1) and (1.4) as

$$\xi_\mu = (\partial/i\partial_\mu)\ln S \quad \text{and} \quad \partial S/i\partial k_\mu = \hat{\xi}_\mu S, \quad (3.3)$$

where $\xi_\mu = (\tau, \bar{\xi})$ is the 4D temporal-spatial function.

The corresponding operator $\hat{\xi}_\mu$ is canonically conjugated with the operator of energy-momentum. So, evidently,

$$i[\hat{\tau}, \hat{H}] = 1 \quad (3.4)$$

and the executing of this obligatory condition underlines the uniqueness of considered operator: it has not place with another artificially introduced duration operators [15]. Note that both operators in (3.4) describe interactions with another objects and with vacuum fluctuations, hence they must act onto S-matrix.

Thus, with these substitutions the Ward-Takahashi identity can be rewritten as

$$k_\mu \xi_\mu(p) = iG(p) k_\mu \Gamma_\mu(p, p', k) \quad (3.5)$$

that at $k_\mu \to 0$ leads to the main definition:

$$\xi_\mu(p) = iG(p)\Gamma_\mu(p, q \to p) \quad (3.6)$$

(it can be deduced directly from the Ward identity, but the used way visually shows that the strong equality k = 0 is not necessary).

Physical sense of duration operator can be précised by the QED representation $\Gamma_\mu(p,q) = \gamma_\mu + \Lambda_\mu(p,q)$ with the Ward identity $\Lambda_\mu(p,p) = -\partial\Sigma(p)/\partial p_\mu$, where $\Sigma(p)$ is the mass operator:

$$\xi_\mu = iG(p)(\gamma_\mu - \partial\Sigma(p)/\partial p_\mu). \quad (3.6')$$

Thus, the durations of scattering process are connected with reorganization of scatterer and with formation of emitted particle. On the other hand (3.6) is close to the expression of current $j_\mu(x) \sim Tr(\Gamma_\mu(x)G(x))$, i.e. each measurement corresponds to the addition of zero energy vertex to corresponding line of the Feynman graph, that justifies the "Larmor clocks" method [12].

The structure of (3.6) shows possibilities of its generalization by the standard substitutions $p \to p - ieA$ with further decomposition over degrees of charge, etc.

Let us consider the principal significance of locality features corresponding to (3.6). The component $\xi_0 \equiv \tau$ describes temporal properties of interaction, other components are related with its spatial lengths, but they can be expressed via the temporal functions and group velocity:

$$\bar{\xi} = i\frac{\partial}{\partial\bar{p}}\ln G(p) = i\frac{\partial}{\partial E}\ln G(p)\frac{\partial E}{\partial\bar{p}} \to -\tau(E)\bar{v}. \quad (3.7)$$

Analogically, for particle of spin ½ from (3.6) follows:

$$\bar{\xi} = iG(p)\bar{\Gamma}(p,q \to p) \to -\tau(E)\bar{\alpha}. \quad (3.8)$$

These expressions allow to writing the covariant representation:

$$\xi_\mu^2 = \tau^2(E)(1 - v^2/c^2), \quad (3.9)$$

i.e. $\xi_\mu$ is the 4-vector of proper duration (cf. [33]).

Note, that axial currents complicate this picture. Instead (3.1) the axial Ward identity ([34], e.g. [35])

$$k_\mu \Gamma_\mu^5(p,q) = \gamma_5 G^{-1}(p) - G^{-1}(q)\gamma_5 - 2im\Gamma^5(p,q) \qquad (3.10)$$

must be written with the axial and pseudo-scalar vertices. It does not lead to such transparent relation for durations as the vector identity (3.1), i.e. will include anomalies. But as can be mentioned, the Ward identities of both types lead to closer results at calculation of particles masses (e.g. [36]). It can be considered as an indication of their equivalence into certain problems, possibly in the determination of durations also (it requires further examinations).

Note that the representation (3.6) may be symmetrized, if needed:

$$\xi_\mu = \frac{i}{2}\{G(p)\Gamma_\mu(p) + \Gamma_\mu(p)G(p)\}. \qquad (3.6'')$$

Vertex part in the local QED is generally represented as

$$\Gamma_\mu(p,q) = \gamma_\mu f(k^2) + i\sigma_{\mu\nu}k_\nu g(k^2) \qquad (3.11)$$

with electric and magnetic form-factors. At substitution of (3.11) into (3.3) the second part falls out, but it can be reconstructed at general decomposition of $\Gamma_\mu$ over Dirac matrices.

The computation of measurable values must be executed with corresponding spinor amplitudes and will be analogical to calculations of transition currents in QED. It leads to sufficiently lengthy expressions and therefore is here omitted. Mention only that the substitution of (3.3) into the dual Schrödinger equation (1.4) shows that if the form-factors are real, the first, electric part is connected with the state formation and the second one with the time-delay at scattering.

The importance of this representation consists in an improvement of the sense of duration of formation: it gives possibilities for establishing distributions descibable by form-factors (3.11). Underline also that the significance of representation (3.6) consists in its evident renormalizability via the known renormalization of its factors.

In the simplest case of electromagnetic interaction with charged particle of zero spin there is only the electric form-factor:

$$\Gamma_\mu^{(s)} \rightarrow (p+q)_\mu f(p-q) \qquad (3.12)$$

and $f(0) = e$ in the nearest approximation over $A_\mu$. It means that scattering processes go without delay and the dressing goes at $G(p) = (p^2 - m^2)^{-1}$ with the duration bigger uncertainty limit.

For particles of spin ½ and with $G(p) = (\gamma p - m)^{-1}$ the first term of (3.11) leads to the expression analogical (3.12) and the second one demonstrates a delay connected with spin. Both terms must include form-factors, for proton and neutron, correspondingly,

$$\Gamma_\mu^{(p)}(p,q) = \gamma_\mu + \frac{1.79}{2m_p}\sigma_{\mu\nu}k_\nu; \qquad \Gamma_\mu^{(n)}(p,q) = -\frac{1.91}{2m_p}\sigma_{\mu\nu}k_\nu, \qquad (3.13)$$

the magnetic terms are expressible through the spin of particle.

The representation (3.6) can be generalized on matrix elements of more complicated structures. If the graph contains (2n+1) lines, among which one photon line can be separated, its feature of transversality leads to the representation:

$$k_\mu M^\mu(k \mid p_1,...,p_n; q_1,...,q_n) = e\sum_l \{M_0(...,q_l - k,...) - M_0(...,p_l + k,...)\}. \qquad (3.14)$$

At $k_\mu \rightarrow 0$ it can be rewritten as

$$M^\mu(k \mid p_1,...,p_n; q_1,...,q_n) \rightarrow -ieM_0(p_1,...,q_n)\sum_l [\xi_\mu^{(l)}(p_l \mid p_1,...,q_n) - \xi_\mu^{(l)}(q_l \mid p_1,...,q_n)],$$

i.e. in the form which includes the temporal characteristics of all lines. Notice, that the described procedure is equivalent to so called ξ-operation used in [37] at consideration of the Pauli-Villars renormalization..

One more feature of temporal functions is related with characteristics of scattering on a system of potentials. Considering amplitudes can be represented as resolvents, i.e. as $1/(1-\gamma K)$, and if $K = K_1+K_2$ then with the identity

$$1 - \gamma K = (1 - \gamma K_1)(1 - \gamma K_2 /(1 - \gamma K_1))$$

follows complete duration via partial ones:

$$\tau = \tau^{(1)} + \tau^{(2)} + \frac{2K_1}{(dK_2/d\omega) - K_1\tau^{(2)}} \tau^{(2)2}, \tag{3.15}$$

this shows a non-additivity, in general, of temporal functions.

\* \* \*

Let's compare these expressions with a mysterious phenomenon of zitterbewegung of Schrödinger [38]: the "momentary" velocity of the free Dirac electron precisely equals $c$. It follows from the time-depended Schrödinger equation with the Hamiltonian $H = \alpha_0 mc^2 + \vec{\alpha}\vec{p}c$ via determinations

$$\partial_t \vec{r}(t) = i[H, \vec{r}] = \vec{\alpha}, \tag{3.16}$$

and

$$\partial_t \alpha_l(t) = i[H, \alpha_l] = 2(i\gamma_l - \sigma_{lj} p_j) = -2i\Gamma_l^{(1)}, \tag{3.17}$$

where the vertex part (3.6) with both form-factors equal to unity is introduced.

This phenomenon is often considered as the evident and unaccountable defect of theory, most often the zitterbewegung of wave packet is attributed, without any further specifications, to fast oscillations of charge, e.g. [39], induced by existence of virtual components with both signs of energy (of charge). The representation (3.9) allows its interpretation via the own spatial length of charge connected with the particle formation.

So, the zitterbewegung occurs within the defined "inner" space of particle. The restriction of oscillations distance reduces a mystery of this phenomenon: it is evidently caused by vacuum fluctuations within the region of own near field and fastly falls down outside of it.

## 4. Light Dispersions

Theory of light dispersion contains some old paradoxes. Let's briefly describe two of them.

1/. The problem of optical forerunners is known for a century: Sommerfeld and Brillouin had shown, within the scope of classical electrodynamics, that the abrupt front part of a light pulse should be passing through any substance without a delay, with vacuum speed c (e.g. [40]). This phenomenon was considered as the consequence of the gradual development of media polarization, necessary, according to the Lorenz-Lorentz formulae for refraction.

In the frame of QED the problem of harbingers is looked differently. Duration of an establishment of polarization is about $\tau_{polar} \sim 1/\omega_0$, duration of its relaxation cannot be bigger time necessary for transfer of sound oscillation on distance $\lambda$: that is $\tau_{relax} \leq \lambda/v_{sound}$. Hence, if the intensity of photon flux $j < j_{min} = 1/\tau_{relax} \lambda^2 = v_{sound}/\lambda^3$, then, according to the theory of polarization development, all photons will be passing through any medium independently, without influence of its features. Numerical value of $j_{min}$ is not very small: in an optical range it corresponds to the light intensity $J_{min}(\omega) = \hbar\omega j_{min} \sim 10^{-4}$ W/cm$^2$, i.e. this effect could be observable in usual conditions as the complete absence of light refraction.

But it would means that isolated photons should go in any media without refraction in general!

2/. Examinations of light pressure on transparent media also have a centenary history, basically as discussions of so called Abraham-Minkowski controversy (e.g. [41]). Two different determinations of classical energy-momentum tensor of electromagnetic field, by Minkowski and by Abraham, lead to different linear momenta of photons in isotropic transparent media: $p_M = n\hbar k/n_g$, $p_A = \hbar k/nn_g$, where $n = c/u$ and $n_g = c/v$ are the phase and group indices of refraction, $u = \omega/k$ and $v = d\omega/dk$ are the phase and group velocities. In accordance with the approach of Minkowski the full momentum is connected to the "photon in medium"; but at the Abraham picture the definite part of this momentum during time of passage is related to the medium. This difference leads to the opposite directions of movement of transparent body at the time of pulse propagation.

This controversy can be presented in the evident paradoxical form. Let's consider the passage of light flux through a transparent medium with the phase velocity $u \leq c$: in conformity with the Maxwell equations, the light momentum in a medium can be bigger than in vacuum and in accordance with the momenta conservation this body should move against a direction of light (Minkowski). But on the other hand such presentation also seems correct: if light is passing through a medium for greater time than in vacuum, the conservation of movement of the center of masses or inertia requires that the body should move in a direction of light flux (Abraham).

These paradoxes evidently show the necessity of reorganization of light propagation theory formerly based on classical electrodynamics, the necessity of the appeal to QED.

\* \* \*

The kinetics of light flux propagation through sufficiently transparent media can be intuitively represented, from the corpuscular point of view, as a sequence of free photons flights with some delays on scatterers, i.e. as a "saltatorial" process instead of continuous waves ones (it allows the consideration of corpuscular optics, influence of external fields, etc.). Such ideas were proposed in the paper [23] and developed with different applications in [24].

In such picture single photon is flying from one scatterer to another with the speed $c$ on the free path length $\ell = 1/\rho\sigma$ ($\rho$ is the density of scatterers, free and valent electrons, $\sigma$ is the cross-section of elastic unbiased scattering). After that it will be detaining on scatterer on the delay time $\tau_1$, determined by (2.14) or by a similar expression for more realistic model. If $\tau_2 > 0$, to the free path length must be added the distance $c\tau_2$, after passing of which photon becomes real, at $\tau_2 < 0$ it instantaneously jumps over the distance $c|\tau_2|$. Thus this photon undergoes $N = L/(\ell+c|\tau_2|)$ acts of elastic unbiased scattering with delays $\tau_1$ by each of them, and therefore the complete duration of its transmission at $\tau_2 > 0$ on the distance $L \gg \ell$ becomes

$$T_+ = L/c + N\tau_1. \tag{4.1}$$

For the case of $\tau_2 < 0$ the substitution $L \rightarrow L - Nc|\tau_2|$ is needed:

$$T_- = L/c + N(\tau_1 - |\tau_2|). \tag{4.1'}$$

Such representation leads to the group refraction index for the region of normal dispersion:

$$n_g = cT/L = 1 + c\tau_1/(\ell + c\tau_2). \tag{4.2}$$

If $\ell \gg c\tau_2$ this expression can be approximately rewritten, sufficiently far from resonances, as

$$n_g \cong 1 + c\rho\tau_1(\omega)\sigma(\omega). \tag{4.2'}$$

These magnitudes allow the estimation of group velocities of photons in transparent media and of corresponding indices of refraction. (Note that, as it was experimentally shown, photon really passes in media with the group velocity [42].)

The phase index of refraction can be determined through the group index as

$$n(\omega) - n_g(0) = \frac{1}{\omega} \int_0^\omega n_g(\omega) d\omega, \qquad (4.3)$$

i.e. by it's averaging over the frequencies interval $(0,\omega)$ at assumption $n_g(0) = 0$ or with a subsequent subtraction.

At the periods of time $N\tau_1$, when the photon is virtually captured, its momentum is transmitted into the medium. But if from the Abraham point of view the part of momentum is transmitted into media onto full duration of photon passage, in the offered approach it takes place at saltatory virtual regime (cf. [43]). If only these times are taken into account, then such part of linear momentum of photon will be related to body:

$$N\tau_1 / T = (n_g - 1)/n_g = 1 - 1/n_g. \qquad (4.4)$$

Hence during the time of photon's pass inside the body, it will possess the momentum

$$p = (1 - 1/n_g)\hbar k \cong \hbar k - \Delta(\hbar k) \qquad (4.5)$$

directed to the flux under $n_g > 1$ and oppositely under $n_g < 1$. It will lead to the displacement of body onto the distance

$$\delta S = pT/M = \pm N\tau_1 \hbar k / M, \qquad (4.6)$$

where $M$ is the mass of displaced body, and as $N \sim L(n_g - 1)/c\tau_1$, the relative displacement of this body at the single photon passage can be expressed as

$$\delta S / L \sim (n_g - 1) \cdot \hbar\omega / Mc^2. \qquad (4.7)$$

Into this expression must be included, generally speaking, displacements caused by the surface effects. This picture, as must be underlined, simultaneously corresponds to both laws of conservations, the linear momenta and the uniform motion of the center of inertia. At the same time it allows the consistent determination of indices of refraction describing the average duration of radiation passage through media.

Notice that the additional momentum in (4.5) is curiously connected with the Abraham and Minkowski momenta:

$$\Delta(\hbar k) = \hbar k / n_g = |p_M \cdot p_A|^{1/2}. \qquad (4.8)$$

At the real absorption by an isolated scatterer (the Bose condensate in [44] must be considered as a single scatterer) the photons' virtual coat, the evanescent waves (see below), must be absorbed together with the energy-momentum (4.5). Therefore at this act, in accordance with conservation laws, $\omega^2 = c^2 k^2$, hence formally $n_g = 1/n$ and therefore

$$\Delta(\hbar k)^{abs} = n\hbar k, \qquad (4.9)$$

i.e. the momentum of body, in accordance with (4.5), will be equal to $p = (1 - n)\hbar k$, just as it is describable by the conception of "photons in medium". So the complete momentum of system after absorption is equal to $\hbar k$, just demands the conservation laws. Such division corresponds to the Minkowski picture, but just as the Abraham picture described above, it is not universally true: both pictures are correct along the definite and different moments and for parts of the total path of photon.

Thus, at the real absorption of single photon by an isolated scatterer of mass M

$$\hbar\omega = (n\hbar k)^2 / 2M + \Delta U , \qquad (4.10)$$

where $\Delta U$ is the energy of inner reconfiguration of scatterer. The term of absorbed kinetic energy can be expressed via the energy of recoil $\hbar\Omega$ and with the explicit accentuation of refraction index factor $\Omega = n^2\Omega'$ it leads to the representation:

$$\hbar\omega = n^2\hbar\Omega' + \Delta U , \qquad (4.10')$$

that conforms to the results of the experiments [44]. Nevertheless, contrary to the conclusions of this article, it does not mean yet that the photon's self-momentum in media is constantly equal to $n\hbar k$: such significance appears at the absorption only, by an integration of parameters of photon self-momentum and of corresponded evanescent waves.

The appearance of evanescent waves propagating mutually with a bare photon can be demonstrated in the frame of QED. The concept of "photons in medium" requires the replacement of the momentum on the pseudo-momentum: $k \to n(\omega)k$. This replacement corresponds to the transformation of causal propagator:

$$D_c(\omega, \vec{k}) \sim \delta_+(\omega^2 - c^2k^2) \longrightarrow D_c(\omega, n(\omega)\vec{k}) \sim \delta_+(\omega^2 - c^2n^2k^2). \qquad (4.11)$$

Inside the medium, far from borders, the accepting of $n = ck/\omega$ is possible. Then by use the properties of $\delta$-function at $\omega \neq 0$ the propagator in transparent isotropic media can be decomposed as

$$D_c(\omega, n\vec{k}) = \tfrac{1}{2}\{D_c(\omega, \vec{k}) + D_c(\omega, i\vec{k})\}. \qquad (4.12)$$

The first term corresponds to the photon's free flight between scatterers with the vacuum speed $c$. The second term of (4.12) describes, accordingly, the field of evanescent waves, caused by photons flight through media; these waves can be attributed to the near field photon's dressing in media: $D_c(\omega, ik) \to D^{(near)}(\omega, k)$. Notice that this term describes the Coulomb field also, inasmuch as $\delta(\omega^2 - c^2k^2) \sim \delta(\omega)\delta(c|\vec{k}|)$ at $\omega \to 0$.

Evanescent waves are characterized by imaginary momenta. As a contrast to the evanescent waves at the FTIR [45] here all components are imaginary. From the formal point of view the Green function $D^{(near)}(\omega, k)$ corresponds to the 4-D Laplace equation for space instantons of zero mass (cf. [46]), it demonstrates that corresponding excitations are transmitting instantaneously.

The difference between two terms of (4.12) is more visual in the mixed ($\omega$, r) representation:

$$D_c(\omega, \vec{r}) = (1/4\pi r)\exp(i|\omega|r/c); \qquad D^{(near)}(\omega, \vec{r}) = (1/4\pi i r)\exp(-|\omega|r/c) \qquad (4.13)$$

that obviously shows the space-like character of near field and its fast attenuation, on account of which near field does not contribute in the far field of sources [31]. Notice that these phenomena on such distances do not overflow the uncertainties limits; therefore the observance of conservations laws is not here required.

Two additional comments.

The described scheme means that passage of photons can be interpreted as the process of sequential measuring of time-delay values. Such measuring, which seems practically impossible at the usual investigations of single scattering processes, becomes possible in optics by virtue of the expression (4.2):

$$\tau(\omega) = \frac{n_{gr} - 1}{cN\sigma_{tot}} \equiv \frac{1}{cN\sigma_{tot}}\left(\frac{d}{d\omega}\omega n(\omega) - 1\right). \qquad (4.14)$$

The basic expression (4.2) can be determined by such way also. If photons are passed through media with the velocity *c* between interactions, and successive interactions are divided by the distances $\ell = 1/N\sigma$, this medium can be considered as a set of planes divided by distances $\ell$, i.e. as the one-dimensional quasi-crystal lattice.

Hence to photon in medium can be attributed any quasi-momentum from the set $k' = k + m\ell$, m=0,1,2,…, i.e. the wave number of photon in medium is determined through the modulo $1/\ell$, and the propagator of photon in transparent medium must be invariable relative to translations onto the vector $\ell$ (an analog of the Bloch theorem for solids). Thereby the construction of an analogue of the Brillouin zone for photons in medium becomes possible at m=1. By substitution the value of free path length and with taking into account that $\tau_1 \approx 1/\omega$ and $k=\omega/c$ for the region of transparency, it can be concluded that

$$k' = k + N\sigma = k(1 + cN\sigma\tau) \equiv n_{ge}(k)k. \tag{4.15}$$

Further considerations with certain applications are given in [24]. Notice that the offered saltatory mechanism of light propagation explains the effect of so called "left turn" of refracted rays in metamaterials [47].

## 5. Multiphoton processes

Let us pass now to multiphoton processes, more concretely to the simplest high harmonics generation (HHG) and multiphoton ionization (MPI) on the bond electron of isolated atom:

$$n\gamma(\omega) + e_B \to e_B + \gamma(\Omega); \qquad n\gamma(\omega) + e_B \to e_{free}(mv^2/2), \tag{5.1}$$

where $n\omega \approx \Omega$, $mv^2/2 \approx \Omega - I$ and $I$ is the potential of ionization. (The method, based on direct calculations of diagrams, was offered in [17], its developments together with several further applications in statistical physics and at problems of astrophysics are described in [18].)

In the process of elastic scattering the electron holds energy of scattered photon during the delay time. If other photons will be scattered during this time on this electron, the virtually held energy can be, correspondingly, increased (duration of holding is determined by the complete captured energy), but will be the most probably emitted at one act, i.e. as the HHG.

Therefore the rate of HHG must be expressed via the density of irradiating flux $j(\omega)$, the delay-time of n photons energy $\tau_1(n\omega)$ and the cross-section of single photon elastic scattering $\sigma(\omega)$. The unique non-dimensional parameter must be composed as $\eta^{(n)} = j(\omega)\sigma(\omega)\tau_1(n\omega)$ and it must be revealed in the scope of QED, i.e. the delay time must be present in theory and must be reflected in its standard expressions.

Matrix element of HHG is of standard form (with possible permutations):

$$M_{fi} = \Psi_f \hat{A}(\Omega) G(n\omega) \hat{A}(\omega) \cdot ... \cdot G(\omega) \hat{A}(\omega) \Psi_i, \tag{5.2}$$

where $G(q\omega) \sim (\omega_0 - q\omega + i\Gamma/2)^{-1}$ (we consider for simplicity two-level non-relativistic systems), and the reaction rate is determined as

$$dR_n = \left| M_{fi} \right|^2 dN_1 \cdot ... \cdot dN_n. \tag{5.3}$$

Each field operator contains the volume of quantization $V^{-1/2}$ and in the limit $V \to \infty$ they lead for in-coming and out-let photons, correspondingly, to the substitutions:

$$\frac{dN_i}{V_i} \to f(t, \vec{r} \mid \omega, k) d\vec{k}; \qquad \frac{1}{V_f} \to \frac{d\vec{K}}{(2\pi)^3}. \tag{5.4}$$

For processes in unidirectional monochromatic fields $f(\vec{k}) \to (j/c)\delta(\vec{k}_0 - \vec{k})$, for processes in heat fields (e.g. at consideration of phase transitions or in astrophysical processes) the Planck distribution must be substituted [18], etc.

Vertex part adds the factor $\Gamma$ to $|G(q\omega)|^2$ and so the time-delay $\tau_1(q\omega)$ is formed. Such manipulations lead to the reaction rate of HHG:

$$R_n^{(HHG)} \sim \left(j\sigma(\omega)\right)^n \Pi_1^n \tau(q\omega), \qquad (5.5)$$

i.e. the temporal functions appear automatically in the course of QED calculations.

This general expression allows consideration of certain case; so, for harmonics of very high numbers $\tau_1(q\omega) \to \Gamma/(q\omega)^2$ and (5.5) is simplified:

$$R_n^{(HHG)} \sim \left(j\sigma(\omega)\Gamma/\omega^2\right)^n /(n!)^2. \qquad (5.6)$$

The consideration of MPI goes analogically, but with a serious complication: the liberated electron must accumulate momentum corresponding to kinetic energy. The time duration of this accumulation can be determined via the Second Law of dynamics, $\tau(\vec{p}) = \Delta\vec{p}/\vec{F}$ with forces on orbit of the ionizable electron. More exact calculations can be executed by the Landau theory of predissociation [48] and its probability is described by the Landau-Zener formula.

Therefore the coordination of two durations, of energy and momentum accumulations, is needed. Note that, in principle, at generation of new states the coordination of durations of several substantial parameters can be demanded.

## 6. Microscopic theory of condensed state: correlation radii, critical indices

Microscopic theories of condensed states must be based on analyses of interactions and bonds between constituents (atoms, molecules): they must be describable, in the frame of QED, via the exchange of virtual photons in the near field and must correspond to the Van der Waals forces.

Emitted photon remains in virtual state during the time $\tau_2$. Therefore the exchange by virtual photons of the certain frequency is possible if the mean free path $\ell$ with the vacuum speed $c$ does not surpass length of the termination (ending) of photon formation. Thus our *basic assumption* can be written down as

$$c|\tau_2(\omega)| \geq \ell(\omega). \qquad (6.1)$$

With this conjecture the natural question arises: what frequencies must be the determinatives in this condition? As formation of the condensed state (gas condensation and/or solidification) are accompanied by the removal of latent heat of phase transitions of the first kind, it seems natural to assume that these frequencies should be proportional just to the latent heat of transition per particle.

Such characteristic frequencies, that can execute bonds between constituents, must be connected to thermodynamic parameters of states and depend on changes at phase transitions. The examination of these bonds can begin with their formation, i.e. with consideration of ways of latent heat removal at entering of each single particle into condensate. In quantum theory this removing the most probably and effectively can be executed via the one-photon emission instead of a gradual calorification of bond energy by each particle connecting to a condensate (symmetries reasons can forbid one-photon transitions, can require two-photons and so on, such features were fixed in several experiments [21]).

The mean free path within a substance $\ell \approx 1/\sigma_{tot}N$, where $\sigma_{tot}$ is the total cross-section of e-γ interactions and $N$ is the density of scatterers (outer electrons). With this definition (6.1) is rewritten as:

$$c|\tau_2|\sigma_{tot}N \equiv V_c(\omega)N \geq 1 \tag{6.1'}$$

The condition of saturation of definite interactions on this frequency corresponds to the equality in (6.1). The offered magnitude $V_c(\omega)$ may be termed 'the volume of interaction'; its magnitude can be generalized by taken into account space-time dependencies, anisotropy of states, polarization effects, external fields, etc. The corresponding frequencies, as we *assume*, are proportional to the (negative) potential energy and, therefore, may be connected with the latent heat, energy of atomization and so on. The non-dimensional magnitude $X = N V_c(\omega)$ or its suitable degree can be considered as the order parameter.

The representation (6.1') with $V_c(\omega) = (4\pi/3)R_c^3$ leads to the effective radius of EM correlations (the length of EM interactions) on this frequency:

$$R_c(\omega) = (3c\tau_2\sigma_{tot}/4\pi)^{1/3} \tag{6.2}$$

In the general case of 3-D spatial interactions, the total cross-section is expressed by the optical theorem of QED as $\sigma_{tot} = (4\pi c/\omega) \cdot \text{Im} A(0)$, where $A(0)$ is the amplitude of unbiased elastic scattering. For low frequencies processes $\text{Im} A(0) \to r_0 = e^2/mc^2$ that leads to the radius of EM correlations or bonds:

$$R_\gamma = (3c^2 r_0)^{1/3} \omega^{-2/3} = 0.28 \cdot 10^{-3} \lambda^{2/3} \text{ [cm]}. \tag{6.3}$$

The condition of saturation in (6.1) can be expressed via the plasma frequency $\omega_P = (4\pi e^2 N/m)^{1/2}$. It allows different interpretations of constituent interactions for corresponding frequency ranges, some of them are considered in [22].

Another interpretation of the interaction volume corresponds to its Fourier transformation:

$$V_\gamma(\vec{r}) = (2\pi)^{-3} \int d\vec{k} \cdot V_\gamma(\vec{k}) \exp(i\vec{k}\vec{r}) = \frac{r_0}{r} = \frac{e^2/r}{mc^2} \tag{6.4}$$

- the interaction volume is determined by the ratio between the bond Coulomb energy and the mass of interacting particle. This expression can be accepted as the basic principle for all subsequent considerations.

With $r \to a_B$, the Bohr radius, it leads to the doubled Rydberg constant. For common distances of the order $r \to 10^8$ cm between atoms/molecules in condensed states, it leads to $R_c \sim 6.8 \times 10^6$ cm, which is of a reasonable order for the long-range interatomic correlations. (Note that such big size of correlations radius in comparison with interatomic distance explains, in particular, the so-called Distler effect [49]: germination of correct crystal structure through the thin metal film that completely covers starting crystal.)

The mole latent energy of phase transitions of the first kind $\Lambda = T(S_2 - S_1)$, where $S_k$ is the entropy of k-th phase. If at approach to the critical temperature $T_c$ this difference smoothly aspires to zero, the latent energy per atom/molecule can be presented as

$$W = \Lambda/N_A \approx T\frac{\partial S}{\partial T}(T_c - T) \to T_c^2\left(\frac{\partial S}{\partial T}\right)_{T=T_c} \Theta \tag{6.5}$$

with $\Theta = |T_c - T|/T_c > 0$.

Let's *suppose* that this energy is distributed between n bonding quanta of equal, for simplicity, frequencies in (6.3): $\hbar\omega \to W/n$. It leads to the radius of virtual photon's correlations:

$$R_c(T) = (3V_c/4\pi)^{1/3} = R_0\,\Theta^{-2/3}. \tag{6.6}$$

Apart from this general bonding of atoms/molecules, the additional interactions caused by the specific parameters of substance constituents could be examined. It can be proposed that these interactions can lead to the specific volumes of interactions, which will provide the features of these substances.

For phase transitions of the second kind the thermal energy of single particle at close range to the critical temperature is determined as

$$W = (1/N_A)\int_T^{T_c} C\,dT \approx \overline{C}T_c\Theta/N_A, \tag{6.5'}$$

i.e. with the similar functional form of correlation radius at $W \to n\hbar\omega$.

Analogical forms may be suggested for other types of interactions via corresponding matrix elements of electrical and magnetic dipole interactions $\hbar\omega \to \vec{d}\vec{E}$ and $\hbar\omega \to \vec{\mu}\vec{H}$. It naturally leads to the radii of correlations:

$$R_{c,d} = R_{0,d}\left(\left|\vec{d}\vec{E}\right|/\left|\vec{d}\vec{E}\right|_{c,d}\right)^{-2/3}; \qquad R_{c,m} = R_{0,m}\left(\left|\vec{\mu}\vec{H}\right|/\left|\vec{\mu}\vec{H}\right|_{c,m}\right)^{-2/3} \tag{6.7}$$

and similar for higher moments. The electron exchanging leads to the correlation radius:

$$R_{c,T} = R_{0,T}\,(T/T_{c,e})^{-2/3}. \tag{6.7'}$$

Hence, all these cases are described by the universal order parameter:

$$R_c \sim \Theta^{-\nu}, \qquad \nu = 2/3, \tag{6.8}$$

where $\Theta \sim \{|T_c - T|, |\mathbf{dE}|, |\mathbf{\mu H}|, T,\ldots\}$ for different types of critical phenomena.

This conclusion supports, in the scope of offered theory, the hypothesis of transitions similarity (the universality of critical behavior, cf. e.g. [50]), all other indices can be now deduced via the known identities.

Let's consider, as an example, the Ginzburg-Landau expansion of thermodynamic potential per unit volume in terms of an order parameter $\eta$ for homogeneous system:

$$\Omega = \Omega_0 + A\eta^2 + B\eta^4 - 2\eta h. \tag{6.9}$$

In their phenomenological theory the simplest order parameter was supposed as $\eta \sim \Theta$, i.e. as the "temperatures distance" till the critical point.

Let us reconsider its possibilities. As the entropy close to transition point is proportional to the temperature "distance" till critical point, $\Delta S \sim \Theta$, then

$$(\Omega - \Omega_0) \sim \Theta^2 \to 1/V_c \tag{6.9'}$$

and it suggests to consider another form of temperature distance till critical point. In accordance with (6.8) and (6.9') the order parameter must be taken as

$$\eta \sim \eta_0\,\Theta^\beta, \qquad \beta = 1/3, \tag{6.10}$$

and the series parameters as

$$A(P, T) = a(P, T)R_c^{-2} = a(P)\mathrm{sgn}(\Theta)R_c^{-2}, \qquad B(P,T) \sim b(P)R_c^{-1}, \tag{6.11}$$

with $A(P, T_c) = 0$.

It can be underlined that experimental data require, in accordance, at least, with the superconductivity results, their representation just in this form [51].

\* \* \*

It seems that the most convincing presentation of the offered conception consists into detailed consideration of the latent heat (bonding energy) removal. As it was firstly proposed in [20], the removal of latent energy at phase transitions of the first kind into more ordered states must be executed by emission of one or certain photons of definite frequencies (such transitions are accompanied, in principle, by the decreasing of symmetry and emission of massless particles in correspondence with the Goldstone or Higgs theories).

This proposition was confirmed by certain special experiments ([21] and references therein) and although there many questions remain, the existence of the effect of radiative removing of latent heat is rigorously established.

## 7. Instantons and Tunneling as Instantaneous Transferring of Excitations

During recent two decades the superluminal transfer of excitations was experimentally observable (speed-faster-c, e.g. the review in [26]). For explanation of these observations we had shown that at strictly definite conditions the "nonlocality in the small" is admissible.

The conditions were formulated as the theorem: "Superluminal transfer of excitations (jumps) through a linear passive substance can be affected by nothing but by the instantaneous tunneling of virtual particles; the tunneling distance is of order of half a wavelength corresponding to the deficiency in the energy relative to the nearest stable (resonance) state. The nonlocality of the electromagnetic field must be described by the 4-potential $A_\mu$, whereas the fields $\vec{E}, \vec{B}$ remain unconnected to the near field".

Its proof in the frame of dispersion relations was presented in [26] and with more details in [27], another proof, via uncertainty principle, is represented in [25], the scheme of such consideration is given above, in the Section 2. This virtual nonlocal formation can be considered as the temporary existence of instanton-like quasi-particle.

Instantons were introduced as the mathematical objects, whose existence is not forbidden[2], but their place in physics, possibilities of their experimental detection are not studied (the history and status of researches e.g. [46]). Instantons correspond to the (formal) substitution $t \to it$ in the relativistic ansatz ($t^2 - r^2$), but such substitution can be real and essential at transition for time $t$ to the duration of interaction $\tau$, when Re$\tau$, the delay of scattering process, is absent and particles (states) continuously are in the formation state. Note that such state can be characteristic for strongly bounded formations, e.g. for quarks.

But we shall consider here more simple problems that can be more easily discussed.

Let us consider the demonstrative and very general WKB example: the 1-D tunneling of particle with energy $E$ through the barrier $U(x)$ described by the quasiclassical matrix element:

$$M \cong \exp\left(-\int_{-a}^{a} dx \sqrt{2m(U(x)-E)}\right), \qquad (7.1)$$

where $a$ is determined from the equality $U(\pm a) = E$.

This representation directly leads to the expression of time durations:

$$\tau(E) \equiv \frac{\partial}{i\partial E} \ln M = -i\sqrt{m/2} \int_{-a}^{a} \frac{dx}{\sqrt{U(x)-E}}. \qquad (7.2)$$

---

[2]). The possible existence of these pseudoparticles once again arouses the famous Wigner's question about the incomprehensible efficiency of mathematics in the natural sciences as the primary problem of the philosophy of science: E.P.Wigner. Comm. Pure and Appl. Math., **13**, 1 (1960).

As the integrand is non-negative at $E < U_0$, we can conclude that $\tau_1=0$, i.e. the tunnel transition is executed without delays and moreover as instantaneous with negative $\tau_2$. For $E > U$ it can be, evidently, $\tau_1 \neq 0$ and both signs of $\tau_2$ are principally possible.

As a more customary example this process can be considered via immediate calculations of wave functions. If on the left side of barrier is the wave packet of sufficiently general form:

$$\psi_i(x,t) = \int_{-\infty}^{\infty} dk \cdot a(k - k_0) e^{i(kx - E(k)t)}, \quad (7.4)$$

which for the Gaussian wave packet and with the approximation $E(k) \to E(k_0) + (k - k_0)v$ can be rewritten as

$$\psi_i(x,t) = A \exp\left(i(k_0 x - E_0 t) - (x - vt)^2 / 2a^2\right), \quad (7.5)$$

then from the standard calculation of transmitted function follows

$$\psi_f(x,t) \cong \psi_i(x - 2a, t) \quad (7.6)$$

(inessential factors are omitted).

This effect of instantaneous transferring of excitation is usually described as an insufficiency of nonrelativistic character of the Schrödinger equation (e.g. [39]). But similar instantaneous bounces, jumps are characteristic for completely relativistic QED expressions also.

Therefore we can think that all tunnel transitions represent instantaneous jumps of excitations. More correctly it can be stated that each tunnel process should be considered as the virtual instanton formation. Hence such old questions as particles movements and some other peculiarities of tunneling lose their sense.

From the general point of view can be concluded that at transition from the variable $t$ to a new variable $\tau$, the ansatz $(t^2 - r^2)$ is substituted onto the ansatz $(\tau^2 - r^2)$. But if $\tau \to i\tau_2$, we have now the Euclidian ansatz - $(\tau^2 + r^2)$, for which the existence of instantons is admissible.

The expression (7.2) shows that the size of instantons must be determined via the minimal energy shortage for completion of free particles formation. It means that any bounded state can be considered via an existence of instantons with parameters determined by the binding energy. In particular, this leads to consideration of condensed states and even bound particles as existing in the peculiar formation states (e.g. the bounded systems in the Coulomb field). Such approach must be, of course, applied to the confinement problems, etc.

Note, that analogical problems arise in many processes that include transition from bare into dressed state and, probably, the most elegant method of solution was offered by Landau in the theory of predissociation [52, 48]. It should be underlined that in his theory this transition is also expressed via the imaginary time, i.e. Landau factually had introduced instantons into theory.

A lot of questions are there raised: how the Euclidian invariance can be connected with the Lorentz invariance, what equations of field may be constructed for near field and so on.

Note that instanton character can be revealed over one of spatial dimensions also. This effect is observable at the photons tunneling, i.e. in the course of so called frustrated total internal reflection, at the phenomenon known from the Newton time and widely used in the theory and practice of waveguides [53].

## 8. Adiabatic hypothesis as implicit explaining of formations duration

For the correct quantum calculation of transition amplitude the procedure of adiabatic incorporating of interaction is needed. Usually it is performed by the substitution $\hat{V}(x) \to \hat{V}(x) \exp(-\lambda|x|)$ for the Hamiltonian of interaction with the subsequent transition $\lambda \to 0$ after performing all calculations, i.e. a temporary narrowing of interaction distances is used. This procedure is known as the adiabatic hypothesis of Ehrenfest: "If a system be affected in a reversible adiabatic way, allowed motions are transformed into allowed motions" [54].

Let's consider the interconnection of this hypothesis with the theory of temporal functions.

The explicit establishment of S-matrix can be carried out by the Bogoliubov method [37] with introducing the covariant switching function $q(x)$, such that $q(x) \to 1$ after performing all calculations. With this purpose S-matrix must be represented as the functional of q:

$$S(q) = P_T \exp\left(i \int dx q(x) L(x.q)\right), \qquad (8.1)$$

$L(x)$ is the Lagrangian of interaction; $P_T$ is the operator of chronologization and the usual 4D designations are used.

*Suppose* that $q(x)$ depends on the type of interaction. As the action function depends on the type of interaction, i.e. on the Lagrangian and also depends on temporal-spatial characteristics of interaction, let us execute the Legendre transformation: $q \leftrightarrow L$. It means that S can be henceforth considerable as the functional from L instead of q:

$$S(L) = P_T \exp\left(i \int dx q(x, L) L(x)\right) = \exp\left(-i \int dk q(-k, L) L(k)\right), \qquad (8.2)$$

where the existence of Fourier transforms of q and L was assumed. Hence a switching of intensity of interaction is replaced onto an (adiabatic) alterations of the interaction 4-volume including, in particular, the duration of interaction, i.e. strictly in the spirit of this hypothesis in quantum mechanics. (Note that the transition to (8.2) can be performed by addition of some classical Lagrangian or source to $L(x, q)$ and subsequent varying over it.)

The variation of (8.2) over L leads to the equations, integral and in variational derivatives:

$$\delta S(L) = i \int dk \cdot q(-k, L) \cdot \delta L(k), \qquad (8.3)$$

$$\frac{\delta S(L)}{i \delta L(k)} = q(-k, L) \cdot S(L). \qquad (8.4)$$

Actually the equation (8.4) already decides the task and shows that the adiabatic factor is directly connected with temporal functions. However, seems interesting to continue this analysis. Let's consider for this purpose the effective Lagrangian $L_1(k)$ with the threshold singularity at $\omega \sim \omega_0$, so that $\partial L_1/\partial \omega$ is a $\delta$-type function at little $|\omega - \omega_0|$ and

$$\delta L_1(k) = \delta_\omega L_1(k) \to (\delta L_1 / \delta \omega); \qquad \delta S(L_1) \sim \delta_\omega S(L_1).$$

With these substitutions in (8.3) the equation

$$\partial S / i \partial \omega = \tau(\omega) S, \qquad (8.5)$$

follows that coincides with (1.4), but with more concretized temporal function:

$$\tau(\omega) = \int dk \cdot q(-k, L_1) \cdot \delta L_1 / \delta \omega. \qquad (8.6)$$

In the covariant form instead of this procedure can be considered a singularity of $L_1(k)$ on a hypersurface $\sigma(k)$, from which follows the equation

$$\delta S(\sigma) / i \delta L(k, \sigma) = q(-k, \sigma) S(\sigma), \qquad (8.4')$$

dual to the completely covariant Tomonaga - Schwinger equation.

In accordance with the adiabatic hypothesis the function q(x) can be represented as

$$q(x) = \exp(-\gamma|t|) \quad \text{or} \quad q(-k) = \delta(\mathbf{k}) \, \gamma/\pi(\omega_0^2 + \gamma^2). \qquad (8.7)$$

The substitution of (8.7) into (8.6) with taking into account the shift $\omega_0 \to \omega - \omega_0$, corresponding to an isolated resonance, leads to the temporal functions used above. The limit transition $\gamma \to 0$ leads to exclusion of all unstable states: at $|t| \to \infty$ every system should contains only the stable states.

The suddenly turning interaction on and off are describable by the switching function:

$$q(x) = \theta(\pm t)\exp(-\gamma|t|), \quad \text{or} \quad q(-k) = \delta(\mathbf{k})\, \gamma/2\pi i(\omega_0 \pm i\gamma)^2 \qquad (8.7')$$

and leads to the temporal functions:

$$\tau = \tau_1 + i\tau_2 = \gamma/\pi[(\omega - \omega_0)^2 + \gamma^2] \pm i(\omega - \omega_0)/\pi[(\omega - \omega_0)^2 + \gamma^2]. \qquad (8.8)$$

Thus it can be concluded that the adiabatic hypothesis presents a non-obvious initiating of the time duration concept in theory

## 9. ON RENORMALIZATION PROCEDURES

Procedures of renormalization are introducing ad hoc, formally, without real physical substantiation. Let us attempt to treat some of these methods with taking into account temporal conception and properties of near field [25].

Infrared divergence is usually eliminated by introduction of fictive photon mass that is directing to zero after all calculations. But this procedure is represented not so formal at taking into account properties of near fields: as was underlined in [31] for their quanta $\omega^2 \neq k^2$ and a "mass of near field photon" has not definite value and sign. Thus, the procedure of transition to zero mass photons means the real, non formal passage from near field excitations to far fields. Such limiting transition has the especial significance and direct physical sense.

The understanding of all particles within near field as virtual excitations with indefinite masses allows consideration of masses bigger than the final (or physical) mass of formatting particle. Such proposition can be considered as a non formal justification of the famous Pauli-Villars method of regularization [55].

The method consists in the substitutions of such type:

$$\Delta(p,m) \to \Delta(p,m) - \Delta(p,M) \sim \frac{m^2 - M^2}{(p - m^2 + i\eta)(p^2 - M^2 + i\eta)} \qquad (9.1)$$

with further passage to the limit $M \to \infty$.

The expression (9.1) leads to the duration of new state formation:

$$\tau_2(p) = \frac{2E}{p^2 - m^2} + \frac{2E}{p^2 - M^2}, \qquad (9.2)$$

i.e. with $p^2 < M^2$ it represents an increasing of the role of more energetic and more deep-seating virtual excitations at the beginning of calculations. Hence this method actually means a partial account of higher terms of S-matrix in the process of particle formation.

Now about the subtraction procedures of renormalization. As a heuristic prompting the existence of derivatives of Green functions over energy can be considered, so the Callan - Symanzik equation of renormalization group contains the operator $\hat{\tau} = \partial/i\partial E$ that is directly related to temporal magnitudes. It must be especially underlined that the method of renormalization group [37] can be reducing directly to the temporal functions. Really, as the corresponding Lie equations contain logarithmic derivatives of propagators over energy-momentum, they are strictly proportional to temporal magnitudes[3].

Let us consider as the example the renormalization of electron propagator. In the course of this procedure the mass function must be represented as

$$\Sigma^{(reg)}(p) = \Sigma(p) - \Sigma(p)|_{\gamma p=m} - (\gamma p - m)\partial_p \Sigma(p)|_{\gamma p=m} \qquad (9.3)$$

with the requirements:

$$\Sigma^{(reg)}(p)|_{\gamma p=m} = 0 \quad \text{and} \quad \partial_p \Sigma^{(reg)}(p)|_{\gamma p=m} = 0 \qquad (9.4)$$

that are usually simply postulated as condition of its renormalization. But they can be interpreted as two physical conditions: first, the mass of free particle has definite magnitude and, secondly, the process of its accumulation to the moment of regularization, at the infinity, is finished.

These requirements must be included in the theory from the beginning as the general physical reasons and therefore the taking of temporal conception into account abolish the formality of this procedure and gives to it the physical sense.

The regularization of the self-energetic part of photon propagator must be physically interpreted as the conditions of the completeness of physical photon formation and the impossibility of its self-acceleration (the details in [25], where several other problems are examined).

Thus the subtraction regularization corresponds to mathematical formulation of the common physical conditions primordially imposed on the system. Therefore these procedures do not represent artificial, ad hoc methods, which usually are formally accepted without attempt to justification of their physical sense.

It can be noted here also that as the dimensional regularization does not violate the Ward identity, the definition of temporal functions can be easily modified for its consideration.

## 10. Duration of interaction and the Landau pole

Let's examine the asymptotic forms of propagators (the approximation of main logarithms) established by Landau, Abrikosov and Khalatnikov in the series of papers summed e.g. in [56]:

$$D_{QED}(k) = \frac{4\pi}{k^2} \cdot \frac{1}{1 + \nu \alpha_c \ln(\Lambda^2/k^2)}, \qquad (10.1)$$

where $\alpha_c = e_c^2/3\pi$, $\nu = 1$ at consideration of vacuum polarization exceptionally by electron-positron pairs and can take into account additional possibilities (here $|k|^2 \gg m^2$, at $|k|^2 \leq m^2$ the substitution $|k|^2 \to m^2$ into logarithm is needed, $\hbar = c = 1$).

In accordance with the Thomson cross-section at $k^2 \to m^2$ the running "coupling constant" $\alpha(k) = (\alpha_c/4\pi)k^2 D(k)$, which leads by substitution of (10.1) to the connection between bare and physical charges (close expressions were established by Gell-Mann and Low also [57]):

$$\alpha_c = \frac{\alpha}{1 - \nu \alpha \ln(\Lambda^2/k^2)}. \qquad (10.2)$$

The expression (10.2) leads to the Landau pole at $\ln(\Lambda_P^2/m^2) = 1/\nu\alpha$, it is the famous problem of charge nullification. The problem is not restricted by QED since (10.1) can be generalized for other field theories at the substitution $\nu \to \beta/4\pi$ and β-functions are calculated for different fields, e.g. [58]. With this generalization and the designation for brevity $\eta_c = \beta\alpha_c/4\pi$ (and below $\eta = \beta\alpha/4\pi$) the propagator summed the main graphs can be rewritten as

$$D(k) = \frac{4\pi}{k^2} \cdot \frac{1}{1 + \eta_c \ln(\Lambda^2/k^2)}. \qquad (10.3)$$

---

[3]. Note that analogical terms are present in certain equations of physical kinetics, but their consideration is far from our aims here.

There are many attempts for suppressing the Landau singularities, e.g. last reviews [59]. But they all can not be recognized as the completed or even sufficiently substantiated: the problem remains intact actuality and requires a search of new approaches.

Let's consider temporal characteristics of these expressions. The duration-length of virtual intermediate boson line is expressed as

$$\xi_\mu \equiv \frac{\partial}{i\partial k_\mu} \ln D(k) = \frac{2ik_\mu}{k^2}\left[1 - \frac{\eta_c}{1 + \eta_c \ln(\Lambda^2/k^2)}\right]. \qquad (10.4)$$

The components of (10.4) give the durations of delay $\tau_1 = \text{Re}\,\xi_0 = 0$ and of formation (dressing):

$$\tau_2 = \text{Im}\,\xi_0 = \frac{2k_0}{k^2}\left[1 - \frac{\eta_c}{1 + \eta_c \ln(\Lambda^2/k^2)}\right] \mapsto \tau_2^{(0)}(1 - \eta), \qquad (10.5)$$

where the first term is related to the uncertainty principle and the second one describes the dynamics of particle formation.

For η>0 this duration is increasing with increasing of $\Lambda \geq m$, i.e. with deepening of interaction, till the value twice bigger the uncertainty limit at $\Lambda \to \infty$. The condition of causality (and analogical for locality) requires the non-negativity of this duration:

$$\tau_2 \geq 0. \qquad (10.6)$$

Therefore the case of 0 < η < 1 does not lead to any complication for $\Lambda \geq m$, i.e. in the domain of used representations. At β ≥ 0 this condition of causality (10.6) leads to the general restriction on the value of running couplings:

$$\alpha(k^2) \leq 4\pi/\beta \qquad (10.7)$$

and as β is increasing with the number of existing fermions in theory and α of strong interaction can be decreased with increasing of energy or vice versa for electromagnetic interactions, this condition restricts, in the principle, the admissible number of fermions in theory.

Let's try to compare these restrictions with the observable data.

In the one-loop approximation with omitting contributions from scalar bosons and so on (e.g. [60]) with suggested existence of only three lepton families $\{\beta_s; \beta_e; \beta_w\} = \{7; 10/3; -4\}$. With the experimental value for strong interaction $\alpha_s = 0.12$ and with the expression $\alpha_e = \alpha(m_Z)\sin^2\theta_W(m_Z)$ for electromagnetic interactions they conform with the restriction (10.7).

However the including of all supersymmetry partners of quarks, leptons, gauge and Higgs bosons (e.g. [60]) leads to the system $\{\beta_s; \beta_e; \beta_w\}' = \{3, -1, -33/5\}$, where the value of $\beta_e$ contradicts (10.7), i.e. such lavish widening overcrowds the system. (The possibilities of optimization of number of additional particles here are not further considered.)

Let us consider the situation when η < 0. It corresponds to conditions: $\eta_c < 0$, i.e. β < 0, and denominator is positive:

$$\ln(\Lambda^2 / k^2) \leq 1 + 1/|\eta_c| \quad \text{or} \quad \Lambda \leq \Lambda_w \equiv |m| \exp\left(\frac{|\eta_c| + 1}{2|\eta_c|}\right). \tag{10.8}$$

Notice that in this case operating momenta do not achieve the Landau pole and its problem becomes non-actual. The common correspondence: $\Lambda / m \propto \lambda_C / R$, where $\lambda_C$ is the Compton wave-length and R is a running distance of field action, allows the estimating of "the radius" of weak interaction:

$$R \geq R_w \equiv \lambda_C \exp\left(-(|\eta_c| + 1)/2|\eta_c|\right). \tag{10.9}$$

With all needed corrections for our estimations $\alpha_w \to 1/171$ and $\beta_w \sim -4$ can be taken. It leads to $|\eta_c|_w = 269$, i.e. $\Lambda_w$ and $R_w$ are close to the inertial mass and to the Compton radius, correspondingly, that allows to consider weak interacting particles as point-like ones.

Note, that the representation of formation duration (10.5) allows an estimation of its alteration with increasing of energy (the Froissart bound [61], standard designation):

$$\sigma_{tot} \leq const \cdot \ln^2 s. \tag{10.10}$$

It leads to the inequality:

$$\tau_2 = \frac{\partial}{\partial E} \ln \sigma_{tot}(E) \leq 2 \frac{\ln s}{s} \cdot \frac{\partial s}{\partial E}, \tag{10.11}$$

and shows the decreasing of formation duration with increasing of energy.

The usage of temporal functions, as can be noted, can resolve also the old discrepancy with non-exponentially of decay [62].

## CONCLUSIONS

Let us sum all considerations and their results.

1. The consecutive theory of the temporal functions describing durations of delay during scattering processes and durations of formation of physical state is examined: the formulas describing these durations are united in the one equation representing the reciprocal analog of the Schrödinger equation.

2. The determination of duration's expressions can be carried out, in particular, on the basis of Ward's identities that introduces them into the context of QED. This form of duration operator can be revealed in well-known QED expressions. It is canonically conjugated with the Hamiltonian that can finish the old known discussions of existence of canonical temporal magnitudes in quantum theory.

3. It allows to demonstrate that the phenomenon of zitterbewegung takes place only in the certain volume determined by the wavelength, it reduces its apparent mysteriousness.

4. Definition of durations of elementary processes allows to examine a kinetics of passage of photons (and other particles also) through substance. It opens possibilities for construction and development of some microscopic (quantum) kinetic theories.

5. Consideration of kinetics of photons passage through substance with taking into account the durations of free run and delays on scatterers leads to calculations of group velocities and phase indices of refraction. This approach resolves, at least, two well known old paradoxes: Sommerfeld-Brillouin and Abraham-Minkowski ones.

6. Consideration of multiphoton processes evidently shows as in the natural way, within the frameworks of QED, temporal functions in the rates of processes appear. They determine thresholds of saturation of processes and opening of new channels of reactions.

7. The condensed substance can be considered as a set of constituents which are in the uncompleted states interacting via near field, i.e. they are not the free, completely terminated particles. This condition allows the determination, in the general form, of the length of electromagnetic correlations between particles depending on energy of interaction (of bond) between them.

8. Such approach confirms (or even proves) the known hypothesis of the universality of critical behavior and enables precise defining of all set of critical indices of phase transitions. In particular, it appears possible to improve a little the coefficients of Ginzburg - Landau equations bringing them into accord with the recent data on superconductivity.

9. Consideration of the phenomenon of tunneling had shown, at least in the quasiclassical approximation, that transfer of excitation through tunnel occurs instantly, i.e. tunneling represents the appearing of virtual instanton. It also executes the phenomenon of «nonlocality in the small» that was established earlier by other methods [26].

10. It is shown that the adiabatic hypothesis of quantum theory represents the implicit introduction of duration of formation in the theory, which corresponds to the switching function of Bogoliubov.

11. The connection between some methods of renormalization in the field theory and the concept of duration of interaction has been shown. It means that renormalization procedures are not formal methods, but their conditions can be initially imposed on the theory

12. The application of this concept to problems of the Landau pole of Green functions leads to a formal restriction on allowable number of fermions in the theory.

In conclusion it can be noted that although, at the first glance, theory of temporal functions seems by an excessive complication (some results were formerly established by well known methods), its generality once more confirms the known maxim of Albert Einstein: *"Make things as simple as possible, but not simpler"*.